\pdfoutput=1

\documentclass[runningheads]{chaos2011}
\usepackage{graphicx,amsmath,amssymb} 
\begin{document}
%
\title*{Transition to Quantum Chaos in Weakly Disordered Graphene Nanoflakes}
%
\toctitle{Transition to Quantum Chaos in Weakly Disordered Graphene Nanoflakes}
\titlerunning{Transition to Quantum Chaos in Graphene}
\author{
  Adam Rycerz
}
\index{Rycerz, A.}

\authorrunning{Rycerz}
\institute{
  Marian Smoluchowski Institute of Physics, Jagiellonian University, Reymonta 4, 30--059 Krak\'{o}w, Poland. 
  (E-mail: {\tt rycerz@th.if.uj.edu.pl})
}

\maketitle             

\begin{abstract}
We analyze numerically ensembles of tight-binding Hamiltonians describing highly-symmetric graphene nanoflakes with weak diagonal disorder induced by random electrostatic potential landscapes. When increasing the disorder strength, statistical distribution of energy levels evolves from Poissonian to Wigner, in\-di\-ca\-ting the transition to quantum chaos. Power laws with the universal exponent map the disorder strength in nanoflakes of different sizes, boundaries, and microscopic disorder types onto a single parameter in additive random-matrix model.
\end{abstract}

\section{Introduction}
Soon after the discovery of graphene---an atomically-thin monolayer of carbon atoms arranged in a honeycomb lattice \cite{Nov04}---it was shown experimentally that electrons in this material behaves as spin--1/2 massless Dirac particles \cite{Nov05}, in agreement with much earlier theoretical prediction by Semenoff \cite{Sem84}. For this reason, the nanostructures in graphene have attracted much attention, leading physicists to reexamine classic effects of quantum transport \cite{Naz09} in search of novel features that arise from the unusual conical band structure, chirality, or the presence of additional quantum number ({\em valley index}) \cite{Bee08,Cas09}. In particular, a Coulomb-blockade experiment on quantum dots consist of graphene nanoflakes and normal metallic leads \cite{Pon08} shown signatures of quantum chaos (the energy-level repulsion) for the flake size smaller then $100\,$nm, but without clear identification of the system symmetry class. Some more light was shed on this issue with theoretical work \cite{Wur09}, showing that measurable quantities may indicate different symmetry class in the case of {\em open} than {\em closed} quantum dot. Later, the energy-level statistics of closed and {\em irregular} graphene flakes obtained from numerical diagonalization of tight-binding Hamiltonians \cite{Lib09,Hua10} was found to coincide with those given by the Gaussian orthogonal ensemble (GOE) of random matrices \cite{Haa10}.

\begin{figure}[t]
  \centerline{
    \includegraphics[width=0.7\textwidth]{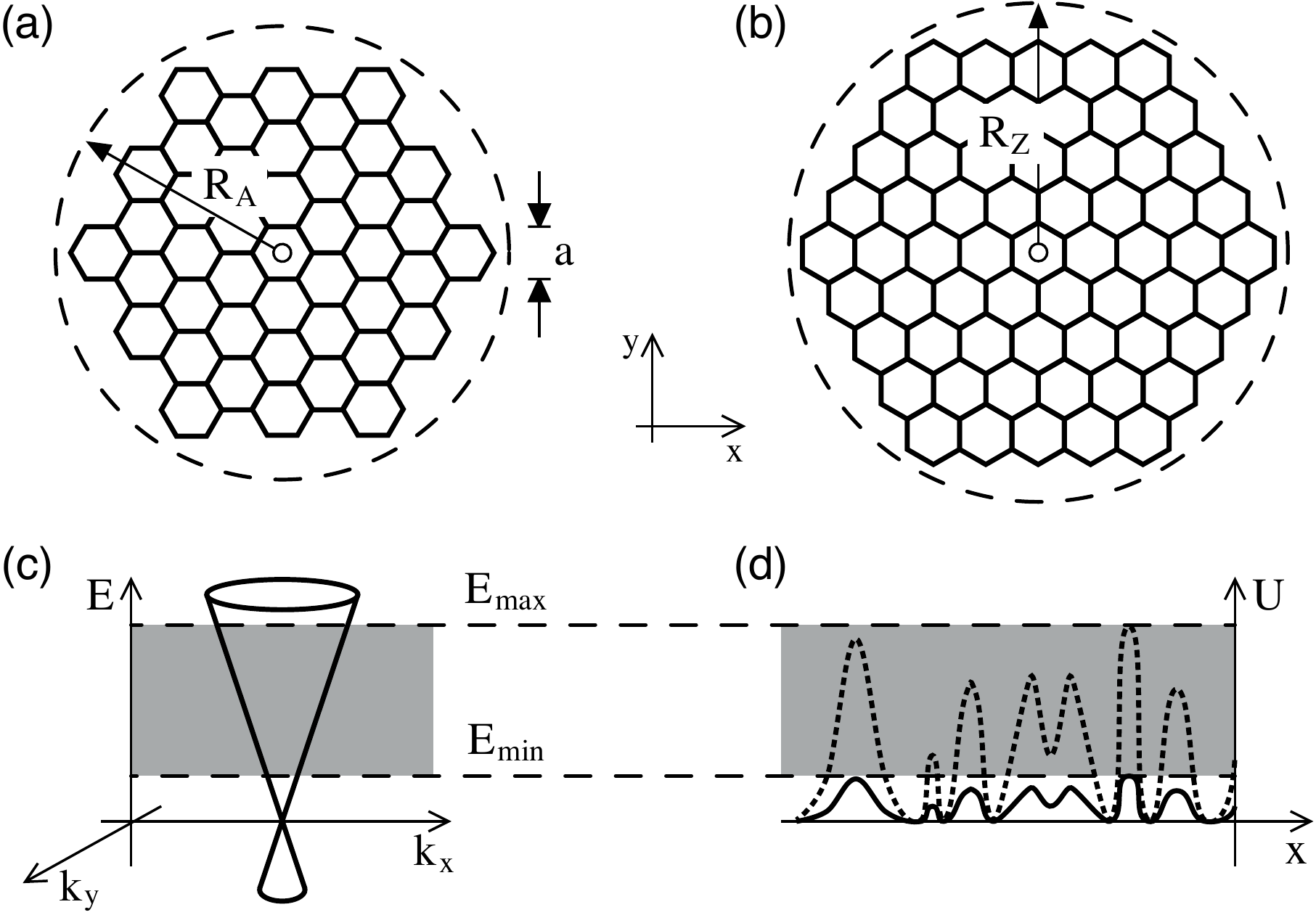}}
  \caption{
    Schematic representation of the systems studied numerically. (a), (b) He\-xa\-go\-nal graphene nanoflakes with armchair and zigzag edges with their radii $R_A$, $R_Z$. (c) Conical dispersion relation $E(k_x,k_y)$ near the Dirac point. (d) Typical potential profiles along the flake. Shaded areas on panels (c), (d) mark the energy range used when discussing the spectral statistics (see the main text for details).}
  \label{fig:hexa}
\end{figure}

In this paper, we follow the numerical approach established by Refs.\ \cite{Wur09,Lib09,Hua10} but focus on {\em regular} (hexagonal) graphene flakes with a weak diagonal disorder attributed to the substrate-induced random electrostatic potential landscape (see Fig.\ \ref{fig:hexa}). The results show, that the energy-level statistics of such systems coincide with those given by additive random matrices of the form $H^0+\lambda{V}$ \cite{Zyc93}, where $H^0$ is the diagonal random matrix (and thus has Poisson statistics) and $V$ is GOE matrix. We also found, that the parameter $\lambda$ is related to the {\em extensive} quantity $N_{\rm tot}K_0$ (where $N_{\rm tot}$ is the total number of carbon atoms and $K_0$ is an {\em intensive} measure of the disorder strength) via the scaling law $\lambda\propto (N_{\rm tot}K_0)^\alpha$, with $\alpha\simeq{}0.6$ regardless boundary conditions and microscopic details of the disorder model.

The paper is organised as follows. In Sec.~\ref{didigra}, we recall the basic findings on possible symmetry classes of chaotic nanosystems containing Dirac fermions, and present microscopic models of disorder in graphene nanoflakes. In Sec.~\ref{ramaspe}, the random matrix model describing the transitions to quantum chaos is applied to rationalize level-spacing distributions obtained from numerical diagonalization of tight-binding Hamiltonians. The conclusions are given in Sec.~\ref{conclu}.

\section{Dirac fermions in disordered graphene\label{didigra}}
In this Section, we present two different microscopic models of disorder in graphene nanoflakes, representing the random electrostatic potential landscape {\em abruptly} or {\em smoothly}  varying on the length-scale of the lattice spacing $a=0.246\,$nm. But first, let us briefly recall (after Ref.\ \cite{Wur09}) the discussion of possible symmetry classes of such nanosystems. 

\subsection{Symmetries of the Hamiltonian}
The effective Hamiltonian for low-energy excitations of electrons in graphene in the absence of magnetic field has a form of the Dirac Hamiltonian
\begin{equation}
\label{hameff}
  {\cal H}_{\rm eff} = 
  v_Fp_x\sigma_x\otimes\tau_z + v_Fp_y\sigma_y\otimes\tau_0
  + \left[M(x,y)\sigma_z + U(x,y)\sigma_0\right]\otimes\tau_0,
\end{equation}
where $v_F\approx{10}^6\,$m/s is the energy-independent Fermi velocity, $\sigma_i$ and $\tau_i$ ($i=1,2,3$) are the Pauli matrices acting on sublattice and valley degrees of freedom (respectively), and $\sigma_0$ ($\tau_0$) denotes the unit matrix. $M(x,y)$ and $U(x,y)$ are the mass term and the external electrostatic potential.
Symmetries of the Hamiltonian (\ref{hameff}) are defined by the following antiunitary operations: standard time reversal ${\cal T}$, and two ``special time reversals''
\begin{equation}
{\cal T}= (\sigma_0\otimes\tau_x){\cal C}, \ \ \ \ 
{\cal T}_{\rm sl}= -i(\sigma_y\otimes\tau_0){\cal C}, \ \ \ \ 
{\cal T}_v= -i(\sigma_0\otimes\tau_y){\cal C},
\end{equation}
where ${\cal C}$ denotes complex conjugation. The mass term breaks the {\em symplectic symmetry} associated with ${\cal T}_{\rm sl}$, leading to the two distinct possible scenarios: 

(i) In the case of weak intervalley scattering, ${\cal T}_v$ commutes with ${\cal H}_{\rm eff}$, so the system consists of two independent subsystems (one for each valley). Each subsystem lacks time-reversal symmetry, as ${\cal T}$ commutes only with full ${\cal H}_{\rm eff}$. Because the Kramer's degeneracy (${\cal T}_v^2=-I$), the Hamiltonian consists of two degenerate blocks, each of which belonging to the Gaussian Unitary Ensemble (GUE). The analogous scenario was considered by Berry and Mondragon \cite{Ber87} for neutrino billiards, lacking the valley degrees of freedom.

(ii) In the case of strong intervalley scattering caused by irregular and abrupt system edges, or by the potential abruptly varying on the scale of atomic separation, the two sublattices are also nonequivalent, so both special time-reversal symmetries ${\cal T}_{\rm sl}$ and ${\cal T}_v$ became irrelevant. ${\cal T}$ commutes with ${\cal H}_{\rm eff}$ leading to the orthogonal symmetry class. 

The existing numerical studies for closed systems of {\em irregular} shapes \cite{Wur09,Lib09,Hua10} show that the typical intervalley scattering time is always shorter than the time required to resolve a level spacing (Heisenberg's time) leading to the scenario (ii). Some features of the scenario (i) were found in open systems \cite{Wur09}, for which the intervalley scattering time needs to be compared with much shorter escape time. Such systems are, however, beyond the scope of this paper, as we focus on {\em regular} and weakly-disordered systems, for which the intervalley scattering itself may be suppressed.

\subsection{Disorder in the tight-binding model of graphene \label{dismod}}
The lattice Hamiltonian for disordered graphene reads
\begin{equation} \label{hamtba}
  {\cal H}=
  \sum_{ij}\gamma_{ij}|{i}\rangle\langle{j}|+
  \sum_i\left[U_{\rm gate}({\bf r}_i)+U_{\rm imp}({\bf r}_i)\right]
  |{i}\rangle\langle{i}|.
\end{equation}
The hopping-matrix element $\gamma_{ij}=-\gamma$ if the orbitals $|i\rangle$ and $|j\rangle$ are nearest neighbors on the honeycomb lattice (with $\gamma=\frac{2}{3}\sqrt{3}\hbar{v}_F/a\approx 3\,$eV), otherwise $\gamma_{ij}=0$. The electrostatic potential contains a contribution $U_{\rm gate}$ from gate electrodes (slowly varying with the site position ${\bf r}_i$) and a random contribution $U_{\rm imp}$ from impurities. For small nanoflakes one can choose $U_{\rm gate}\simeq{}U_0=0$, whereas a realization of disorder potential is generated by randomly choosing $N_{\rm imp}$ lattice sites ${\bf R}_n$ ($n=1,\dots,N_{\rm imp}$) out of $N_{\rm tot}$, and by randomly choosing the amplitudes $U_n\in(-\delta,\delta)$. The potential is then smoothed over a distance $\xi$ by convolution with a Gaussian, namely
\begin{equation} \label{uimper}
  U_{\rm imp}({\bf r})=
  \sum_{n=1}^{N_{\rm imp}}U_n\exp\left(-\frac{|{\bf r}-{\bf R}_n|^2}{2\xi^2}\right).
\end{equation}
The special case of $\xi\ll{a}$, $N_{\rm imp}=N_{\rm tot}$ corresponds to the Anderson model on a honeycomb lattice, considered in work \cite{Ama09} on spectral statistics of nanotube-like structures. Earlier, the model constituted by Eqs.\ (\ref{hamtba},\ref{uimper}) with $\xi\gg{a}$ was shown to reproduce basic transport properties of disordered mesoscopic graphene samples \cite{Ryc07,Lew08}. It has not been considered, however, in the discussion of spectral statistics of nanoflakes so far.

We further define the Fourier transform of two-point correlation function
\begin{equation} \label{knoddef}
  K_{\bf q}=\frac{\cal A}{\left(N_{\rm tot}\hbar{v}_F\right)^2}\sum_{i=1}^{N_{\rm tot}}
  \sum_{j=1}^{N_{\rm tot}}\left\langle
    {U_{\rm imp}({\bf r}_i)U_{\rm imp}({\bf r}_i)}
  \right\rangle\exp\left[i{\bf q}\cdot({\bf r}_i-{\bf r}_j)\right],
\end{equation}
where the system area ${\cal A}=\frac{1}{4}\sqrt{3}N_{\rm tot}a^2$, and the averaging takes place over possible realizations of the disorder (\ref{uimper}) (so $\langle{U_{\rm imp}}\rangle=0$). For the length scales large compared to $\xi$, the dimensionless correlator
\begin{equation} \label{knodval}
  K_0=\frac{\sqrt{3}}{9}\frac{N_{\rm imp}}{N_{\rm tot}}
  \left(\frac{\delta}{\gamma}\right)^2\kappa^2, \ \ \ \ \ \ 
  \kappa=\begin{cases}
    1, & \text{if } \xi\ll{a}, \\
    \frac{8}{3}\sqrt{3}\pi(\xi/a)^2, & \text{if } \xi\gg{a},
  \end{cases}
\end{equation}
becomes a representative measure of the disorder strength. For ${\bf q}\neq{0}$,
we obtain $K_{\bf q}=K_0$ if $\xi\ll{a}$, or $K_{\bf q}=K_0\exp(-q^2\xi^2)$ if $\xi\gg{a}$. The numerical value of the ratio $K_{\bf q}/K_0$ at ${\bf q}=\left(\pm\frac{2\pi}{3a},0\right)$ approximates the intervalley scattering rate, and is as small as $2\times{10}^{-6}$ for $\xi=\sqrt{3}\,a$ (used in the numerical simulations presented in remaining parts of the paper).

\section{Random matrices and spectral statistics\label{ramaspe}}
\subsection{Additive matrix model for transition Poisson-GOE}

Before presenting the numerical results for spectral statistics of graphene nanoflakes, let us briefly review corresponding additive random-matrix models and resulting nearest-neighbor spacings distributions \cite{Zyc93}. 

When large integrable system undergoes transition to quantum chaos, its spectral properties can be modelled by the following random Hamiltonian
\begin{equation}
  \label{admatm}
  H=\frac{H^0+\lambda{V}}{\sqrt{1+\lambda^2}},
\end{equation}
where $H^0$ is diagonal random matrix, which elements follow a Gaussian distribution with zero mean and the variance $\langle(H^0_{ij})^2\rangle=\delta_{ij}$, the parameter $\lambda\in{}[\,0,\infty\,]$, and $V$ is a member of one of the Gaussian ensembles. In particular, for transition Poisson-GOE, elements of $V$ are real numbers chosen to follow a Gaussian distribution with zero mean and the variance $\langle{V_{ij}^2}\rangle=(1+\delta_{ij})/N$, where $N$ is the matrix size.

For $N=2$, the eigenvalue-spacings distribution for the Hamiltonian (\ref{admatm}) can be found analytically and reads, for transition Poisson-GOE,
\begin{equation}
  \label{pspgoe}
  P(\lambda;S)=
  \left[\frac{u(\lambda)^2S}{\lambda}\right]
  \exp\left[{-\frac{u(\lambda)^2S^2}{4\lambda^2}}\right]
  \int_0^\infty\!{d\eta}e^{(-\eta^2-2\lambda\eta)}
  I_0\left[\frac{\eta{u(\lambda)}S}{\lambda}\right].
\end{equation}
$I_0(x)$ is the modified Bessel function of the first kind; $u(\lambda)=\sqrt{\pi}U(-\frac{1}{2},0,\lambda^2)$ with $U(a,b,x)$ the confluent hypergeometric function \cite{Abram}. In particular, for $\lambda=0$ the Poissonian distribution $P(S)=\exp(-S)$ is restored. For the opposite limit ($\lambda\rightarrow\infty$) we have $P(S)=(\pi/2)S\exp(-\pi{S^2}/4)$, reproducing the Wigner surmise for GOE matrices. For $0<\lambda<\infty$, Eq.\ (\ref{pspgoe}) describe level-spacings distributions interpolating between Poisson and GOE statistics, with $P(\lambda;S)\propto{S/\lambda}$ if $S\lesssim\lambda\ll{1}$, or $P(\lambda;S)\propto{S}$ if $S\ll{1}\lesssim\lambda$.

For large $N$, the statistics $P(S)$ (so-called {\em nearest-neighbor spacings distribution}) is defined as a distribution of a variable $S=(E_{n+1}-E_n)\langle\rho\rangle$, where $\langle\rho\rangle$ is the average density of states, and $E_n<E_{n+1}$ are neighboring energy levels. Subsequently, we have $\int_0^\infty{P(S)}=\int_0^\infty{SP(S)}=1$ (so-called {\em unfolded spectrum}). Although Eq.\ (\ref{pspgoe}) is exact for $N=2$ only, it was shown numerically \cite{Zyc93} that $P(\lambda_{\rm fit};S)$ with $\lambda_{\rm fit}\simeq\sqrt{N}\lambda$ provides an excellent approximation of $P(S)$ for large random matrices of the form given by Eq.\ (\ref{admatm}).

\begin{figure}[t]
  \centerline{
    \includegraphics[width=0.8\textwidth]{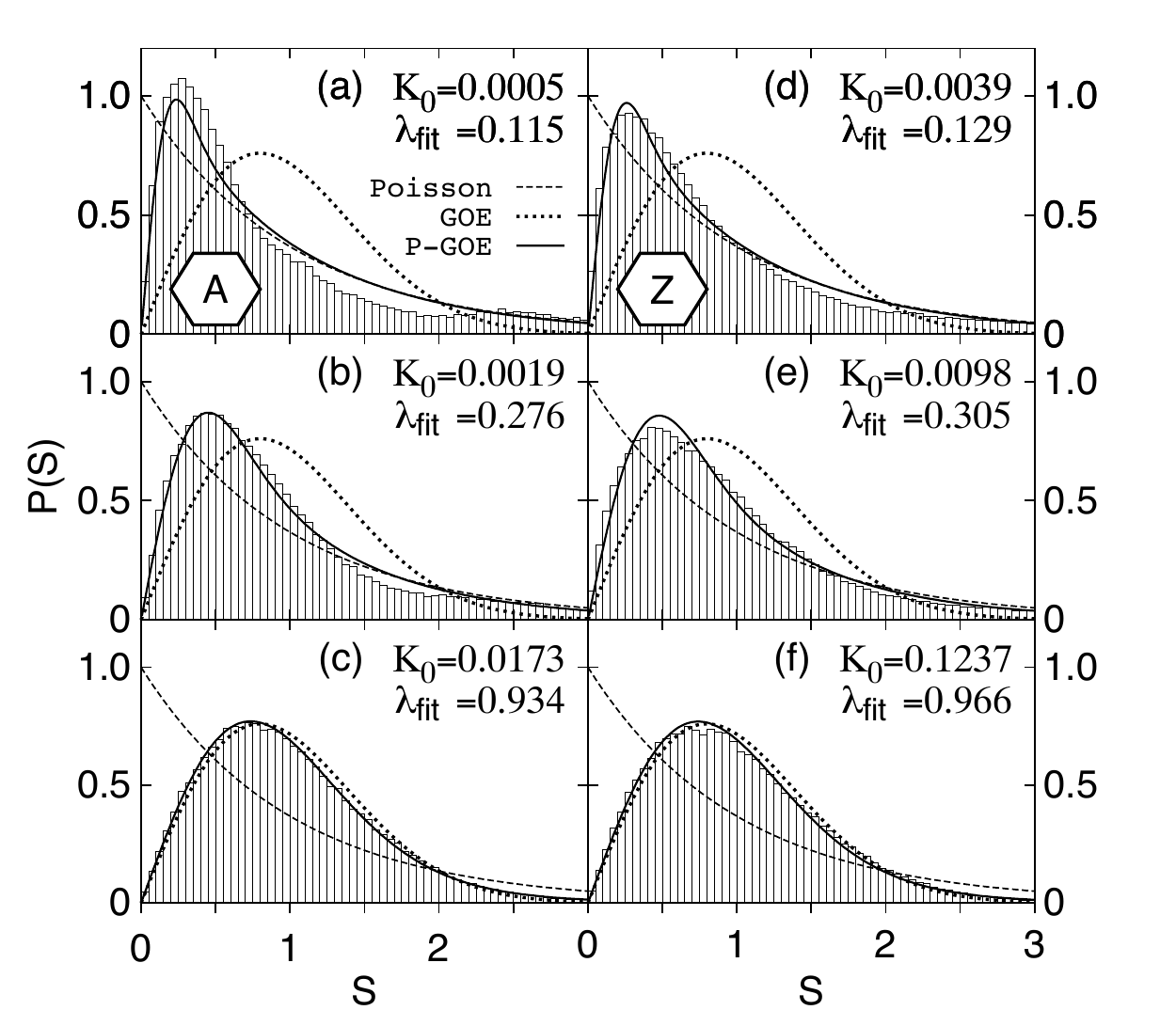}}
  \caption{
    Nearest-neighbor spacing distribution $P(S)$ for hexagonal flakes of Fig.\ \ref{fig:hexa}. (a)--(c) Armchair edges, Anderson model of disorder ($\xi=0$, $N_{\rm imp}=N_{\rm tot}=8322$). (d)--(f) Zigzag edges, smooth impurity potential ($\xi=\sqrt{3}a$, $N_{\rm imp}\ll{N}_{\rm tot}=10584$). Disorder strength $K_0$ (\ref{knodval}) is varied between the panels by changing the potential amplitude $\delta$ [panels (a)--(c)] or by fixing $\delta/\gamma=0.1$ and varying the impurity concentration $N_{\rm imp}/N_{\rm tot}$ [panels (d)--(f)]. Histograms show the numerical data obtained by averaging over 200--400 disorder realizations. Solid lines show the statistics interpolating between Poisson and GOE (\ref{pspgoe}) with best-fitted parameter $\lambda=\lambda_{\rm fit}$ specified for each panel. The limiting cases of Poisson ($\lambda=0$) and GOE ($\lambda=\infty$) statistics are shown with dashed and dotted lines (respectively). }
  \label{fig:psazh}
\end{figure}

\begin{figure}[t]
  \centerline{
    \includegraphics[width=0.7\textwidth]{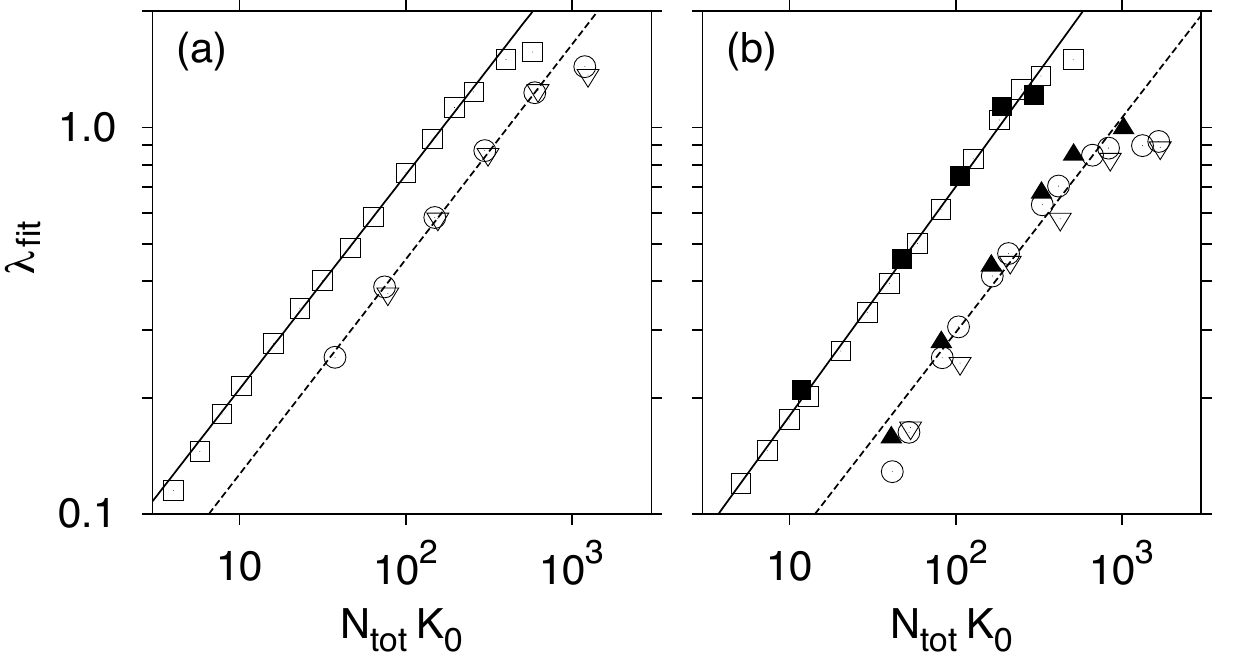}}
  \caption{
    Least-squares fitted parameters $\lambda_{\rm fit}$ for transition Poisson-GOE (\ref{pspgoe}) as functions of disorder strength for hexagons with armchair edges (a) [or zigzag edges~(b)], different sizes, and the two distinct disorder types. $N_{\rm tot}=8322$~(a) [or $10584$~(b)]: $\xi=0$ ($\square$), $\xi=\sqrt{3}a$ ($\bigcirc$); $N_{\rm tot}=34062$~(a) [or $42366$~(b)]: $\xi=\sqrt{3}a$ ($\bigtriangledown$); $N_{\rm tot}=6144$~[panel (b) only]: $\xi=0$ ($\blacksquare$), $\xi=\sqrt{3}a$ ($\blacktriangle$). Lines denote best fitted power-law relations for the two disorder types (see Table~\ref{tab:lbfit} for details). }
  \label{fig:lbfit}
\end{figure}

\subsection{Energy-level distributions for disordered graphene flakes}
In this Subsection, the central question of the present work is addressed, namely: Whether the statistic interpolating between Poisson and GOE, $P(\lambda;S)$ (\ref{pspgoe}) is capable of describing nearest-neighbor spacings distributions $P(S)$ for weakly-disordered graphene flakes? In other words, {\em may the additive-random matrix model defined via Eq.\ (\ref{admatm}) be applicable for such relativistic nanosystems?} To answer this question, we focus on two systems of a high symmetry: hexagonal flakes with entirely armchair or zigzag edges, each of which is showing Poisson statistic in the absence of disorder (providing the level degeneracy is properly taken into account). As already mentioned in Sec.\ \ref{didigra}, two distinct models of disorder are applied to each system: {\em Anderson model}, defined by setting $\xi=0$ and $N_{\rm imp}=N_{\rm tot}$ in Eqs.\ (\ref{hamtba},\ref{uimper}), or {\em smooth disorder}, with $\xi=\sqrt{3}a$ and $N_{\rm imp}\ll N_{\rm tot}$. 

To obtain the statistics $P(S)$, we diagonalized numerically tight-binding Hamiltonians (\ref{hamtba}) for the flake containing $N_{\rm tot}\lesssim{10^4}$ atoms and $200-400$ independent disorder realizations for either type of edges, disorder models, and each disorder strength quantified by the correlator $K_0$ (\ref{knodval}). Some additional effort is required when unfolding the spectra: Unlike for two-dimensional gas of Schr\"{o}dinger electrons, for which average density of states $\langle\rho\rangle$ is assumed to be energy-independent, for bulk graphene we have \cite{Ber87} $\langle\rho(E)\rangle\simeq{\cal A}|E|/[\pi{}(\hbar{v_F})^2]$ (per spin). For small systems studied here, boundary effects lead to additional states appearing near $E\simeq\pm\gamma$ (armchair edges) or $E\simeq{0}$ (zigzag edges). Also, the impurity potential (\ref{uimper}) introduces some bound states for $|E|<\delta$. All these additional states, however, are localized on areas small in comparison to ${\cal A}$, and thus not contribute to the spectrum obtained in a Coulomb-blockade experiment such as reported in Ref.\ \cite{Pon08}. For this reason, we limit the energy range [cf.\ $E_{\rm min}$ and $E_{\rm max}$ in Fig.\ \ref{fig:hexa}(c),(d)] such that 
\begin{equation}
  \label{rhoeff}
  \langle\rho(E)\rangle\simeq\rho_0+\frac{1}{\pi}\frac{{\cal A}_{\rm eff}}{(\hbar{}v_F)^2}|E|,
  \ \ \ \  \mbox{for}\ \ 0.1\leqslant{}|E|/\gamma\leqslant{}0.5.
\end{equation}
The constant term $\rho_0$ and the effective area ${\cal A}_{\rm eff}\lesssim{\cal A}$ are determined via least-square fitting of Eq.\ (\ref{rhoeff}) to the actual $\langle\rho(E)\rangle$ obtained by numerical averaging over independent disorder realizations.

Our numerical results are presented in Figs.\ \ref{fig:psazh} and \ref{fig:lbfit}. First, we compare the statistics $P(S)$ on two selected examples of nanosystems considered: the hexagon with armchair edges and Anderson-type disorder (Fig.\ \ref{fig:psazh}(a)--(c)) and the hexagon with zigzag edges and smooth disorder (Fig.\ \ref{fig:psazh}(d)--(f)). Although some systematic deviations of $P(S)$ from the best-fitted interpolating statistics $P(\lambda_{\rm fit};S)$ (\ref{pspgoe}) are visible for $S>{1}$ due to a~finite system size (notice that a better agreement is observed for $N_{\rm tot}=10584$ than for $8322$), $P(\lambda_{\rm fit};S)$ reproduces the actual nearest-neighbor spacings distribution with a good accuracy for both systems and wide range of $K_0$. We further notice, that similar values of $\lambda_{\rm fit}$ are reached for the second system at $K_0$ typically $5-8$ times larger than for the first system.

\begin{table}[t]
  \caption{ \label{tab:lbfit}
    Least-square fitted power-laws $\overline{\lambda}_{\rm fit}(\zeta)=
    \lambda_1\zeta^\alpha$, with $\zeta\equiv{N_{\rm tot}}K_0$ 
    (lines in Fig.\ \ref{fig:lbfit}). 
    Numbers in parenthesis are standard deviations for the last digit.}
  \centerline{
    \begin{tabular}{ll|rr|rr} \hline\hline
      \multicolumn{2}{c}{\bf Disorder model} 
      & \multicolumn{2}{|c|}{\sf Armchair edges} 
      & \multicolumn{2}{c}{\sf Zigzag edges} \\ 
      \hline
      $\ \xi\!=0$, & $\ N_{\rm imp}\!=N_{\rm tot}\,$ 
      & $\ \lambda_1\!=0.059(2)\,$ & $\ \alpha\!={0.55(1)}\,$ 
      & $\ \lambda_1\!=0.046(3)\,$ & $\ \alpha\!={0.59(1)}\ $ \\
      $\ \xi\!=\sqrt{3}a$, & $\ N_{\rm imp}\!\ll{}N_{\rm tot}\,$ 
      & $0.035(2)\,$ & ${0.56(1)}\,$ 
      & $0.023(4)\,$ & ${0.56(3)}\ $ \\ \hline\hline
    \end{tabular}
  }
\end{table}

The dependence of $\lambda_{\rm fit}$ on the {\em total disorder strength} $N_{\rm tot}K_0$ for all datasets available is illustrated in Fig.\ \ref{fig:lbfit} (datapoints) in the logarithmic scale. The particular choice of the independent variable $\zeta\equiv{N_{\rm tot}K_0}$ allows us to find the approximating relations $\lambda_{\rm fit}\simeq\overline{\lambda}_{\rm fit}(\zeta)$, which still differ between the systems with different edges or disorder types, but remain unchanged when varying $N_{\rm tot}$ and $K_0$ independently with the remaining parameters fixed. Also, for smooth disorder, we vary $N_{\rm imp}$ having $\delta$ fixed at $\delta/\gamma=0.1$ or $0.5$ (corresponding to the absence or presence of charge puddles in the physical system).

Least-square fitted power-laws $\overline{\lambda}_{\rm fit}(\zeta)$ are listed in Table \ref{tab:lbfit} and plot in  Fig.\ \ref{fig:lbfit}(a),(b) (lines). The power-laws fail for $\lambda_{\rm fit}\gtrsim{1}$, as $P(\lambda_{\rm fit};S)$ becomes indistinguishable from GOE statistics in this range. They are, however, closely-followed by the datapoints for smaller  $\lambda_{\rm fit}$-s. We further verify the obtained $\overline{\lambda}_{\rm fit}(\zeta)$-s for the case of smooth impurity potential, by taking the systems approximately four times larger in area (namely, $N_{\rm tot}=34062$ for armchair edges or $42366$ for zigzag edges), but generating only one disorder realization for each $K_0$. Such an approach reproduces the experimental procedure of Ref.\ \cite{Pon08}, where the spectrum of a single system was obtained. Additionally, the corresponding flake diameters $2R_A=87\sqrt{3}\,a\simeq{}37\,$nm and $2R_Z=168\,a\simeq{}41\,$nm are of the same order of magnitude as diameters reported in Ref.\ \cite{Pon08}. The new datapoints (open triangles in Fig.\ \ref{fig:lbfit}) still follow the corresponding power-laws, providing that $\lambda_{\rm fit}\lesssim{1}$.

Probably, the most remarkable feature of these results is that all graphene nanoflakes considered show transition Poisson-GOE when increasing the disorder strength, with no signatures of GUE statistics. This is expected for the flakes with armchair edges which couple the valleys \cite{Bee08}, or with zigzag edges and Anderson-type disorder \cite{Ryc07}, for which the intervalley scattering restores time-reversal symmetry. The absence of GUE statistics seems surprising in the case of zigzag edges accompanied by the smooth impurity potential. In such case, some intervalley scattering originates from six $120^\circ$ corners, a role of which may become decisive for spectral statistics of closed nanosystems.

\section{Conclusions \label{conclu}}
We find that the additive random-matrix model, describing a transition to quantum chaos in Hamiltonian systems, is also relevant when discussing spectral statistics of highly-symmetric graphene nanoflakes with a weak diagonal disorder. The functional relation between the model parameter $\lambda$ and the disorder strength $N_{\rm tot}K_0$ has a form of a power law, with the universal exponent $\alpha\simeq{0.6}$, which is insensitive to the boundary type or to the microscopic model of the impurity potential. 

In the chaotic range, \emph{regular} graphene flakes show energy-level statistics characteristic for the Gaussian Orthogonal Ensemble (GOE) of random matrices, indicating the strong scattering of Dirac fermions between the valleys. This coincides with earlier findings for \emph{irregular} nanoflakes \cite{Wur09,Lib09,Hua10}.

\section*{Acknowledgments}
The work was supported by the National Science Centre of Poland (NCN) via Grant No.\ N--N202--031440, by the Alexander von Humboldt Stiftung-Foundation (AvH), and partly by Foundation for Polish Science (FNP) under the program TEAM.

\end{document}